\documentclass[journal]{IEEEtran}
\usepackage[keeplastbox]{flushend}

\ifCLASSINFOpdf
\usepackage[pdftex]{graphicx}
\graphicspath{./imgs/}
\DeclareGraphicsExtensions{.pdf,.jpeg,.png}
\else
\usepackage[dvips]{graphicx}
\graphicspath{./imgs/}
\DeclareGraphicsExtensions{.eps}
\fi

\usepackage{epstopdf}

\usepackage[cmex10]{amsmath}
\usepackage{mathtools} 
\usepackage{color}

\ifCLASSOPTIONcompsoc
\usepackage[caption=false,font=normalsize,labelfont=sf,textfont=sf,justification=centering]{subfig}
\else
\usepackage[caption=false,font=footnotesize,justification=centering]{subfig}
\fi

\usepackage{stfloats}
\usepackage[hyphens]{url}

\newcommand{\f}[1]{\ensuremath{\mathbf{#1}}} 
\newcommand{\fg}[1]{\ensuremath{{\boldsymbol{#1}}}} 
\newcommand{\mat}[1]{\ensuremath{\boldsymbol{#1}}} 

\def\R{\mathbb{R}} 
\def\C{\mathbb{C}}
\def\Z{\mathbb{Z}}

\def\TVE{\textrm{TVE}}

\def\FE{\textrm{FE}}
\def\RFE{\textrm{RFE}}
\def\DE{\textrm{DE}}
\def\RDE{\textrm{RDE}}

\def\STA{\textrm{STA}}

\def\SNR{\textrm{SNR}}

\def\diag{\textrm{diag}}
\def\r{\textrm{pr}}
\def\con{\textrm{con}}
\def\res{\textrm{res}}

\renewcommand*{\Re}{\operatorname{Re}}
\renewcommand*{\Im}{\operatorname{Im}}

\usepackage{algorithm}
\usepackage{algpseudocode}
\usepackage{amsfonts}
\usepackage{widetext}
\usepackage{float}
\usepackage{amssymb}
\usepackage[english]{babel}
\usepackage{multirow}
\usepackage{mathdots}
\usepackage[noadjust]{cite}
\usepackage{siunitx}

\algnewcommand\algorithmicinput{\textbf{Input:}}
\algnewcommand\Input{\item[\algorithmicinput]}
\algnewcommand\algorithmicoutput{\textbf{Output:}}
\algnewcommand\Output{\item[\algorithmicoutput]}
\newcommand{\pushcode}[1][1]{\hskip\dimexpr#1\algorithmicindent\relax}

\begin{document}
%



\title{\color{black}{A Self-Adaptive Contractive Algorithm for Enhanced Dynamic Phasor Estimation}}

%
%




\author{Francisco Messina, Pablo Marchi, Leonardo Rey Vega, Cecilia G. Galarza\thanks{This work was supported by grant UBACyT 20020170100470BA from the University of Buenos Aires. F. Messina is with the Department of Electrical and Computer Engineering, McGill University, Montreal, Quebec, Canada (e-mail: francisco.messina@mail.mcgill.ca). P. Marchi, L. Rey Vega and C. G. Galarza are with the CSC-CONICET and the School of Engineering - Universidad de Buenos Aires, Buenos Aires, Argentina.}}

\maketitle

\begin{abstract}

In this paper, a self-adaptive contractive (SAC) algorithm is proposed for enhanced dynamic phasor estimation in the diverse operating conditions of modern power systems. At a high-level, the method is composed of three stages: parameter shifting, filtering and parameter unshifting. The goal of the first stage is to transform the input signal phasor so that it is approximately mapped to nominal conditions. The second stage provides estimates of the phasor, frequency, rate of change of frequency (ROCOF), damping and rate of change of damping (ROCOD) of the parameter shifted phasor by using a differentiator filter bank (DFB). The final stage recovers the original signal phasor parameters while rejecting misleading estimates. The most important features of the algorithm are that it offers convergence guarantees in a set of desired conditions, and also great harmonic rejection. Numerical examples, including the IEEE C37.118.1 standard tests with realistic noise levels, as well as fault conditions, validate the proposed algorithm.


\end{abstract}

\begin{IEEEkeywords} 

Adaptive systems, contractive algorithm, differentiator filter banks, dynamic phasor estimation, harmonic rejection, phasor measurement units, rate of change of damping, semi-infinite optimization, wide range of conditions.

\end{IEEEkeywords}

%
\IEEEpeerreviewmaketitle


\section{Introduction} \label{sec:introduction}
%
%
%
%

\IEEEPARstart{T}{he} proliferation of nonlinear loads and incorporation of distributed energy resources (DERs) are dramatically changing modern power systems \cite{ipakchi2009}. In particular, their typically large inertia is steadily decreasing \cite{tielens2016}, which means that they are operating in more varied conditions, making them more prone to disturbances. Under this scenario, monitoring, protection and control of power systems are becoming much more challenging problems. However, the increasing deployment of phasor measurement units (PMUs) is allowing the development of wide-area measurement systems (WAMS), which in turn provides the opportunity to enhance the grid stability assessment, state estimation, fault detection, generator model validation, etc. \cite{aminifar2014}.

Phasor estimation accuracy has an impact on all the aforementioned applications and is, therefore, a problem of fundamental importance in modern power systems. Although the IEEE Std.  C37.118.1-2011 \cite{ieeestd2011} on synchrophasors and its recent amendment C37.118.1a-2014 \cite{ieeestd2014} (from now on simply denoted as the IEEE Std.) are firmly adopted for evaluation and certification of PMUs, it has received criticism, some of which are the following. First, given the importance of phase estimation accuracy in several applications, including active power flow computations and grid restoration \cite{phadke2017}, phase error (PE) may be a more useful metric than total vector error (TVE), a joint measure of the amplitude and phase errors \cite{barchi2015}. On the other hand, amplitude error (AE) may be critical for other cases, such as high-impedance fault detection and voltage sag analysis \cite{meier2017}. Therefore, besides the TVE, it is important to characterize AE and PE separately. Another criticism arises from the modification and suspension of some of the RFE limits in the harmonic and interharmonic interference tests, since this could seriously limit the application of PMUs in island detection schemes and generation control for grid stabilization \cite{roscoe2015}. Similarly, since islanded grid operation may produce temporary frequency offsets even higher than those stipulated in the IEEE Std. \cite{phadke2017}, there is a current interest in wide frequency range algorithms \cite{kamwa2014,zhan2018}. In addition, the IEEE Std. does not consider the impact of acquisition noise but this effect may be significant, particularly for frequency and rate of change of frequency (ROCOF) measurements \cite{macii2016}. On the other hand, in distribution systems the angular differences between nodes are quite small, so much more accuracy is typically required \cite{meier2017}. From the previous discussion, one could argue that the complete specification of PMUs performance requirements in different scenarios and applications is still an open problem. In any case, it is clear that next-generation PMUs should aim at providing enhanced performance.

In this context, a considerable amount of research was conducted in synchrophasor estimation algorithms \cite{aminifar2014}, and interest in this area is still growing. These algorithms may be classified broadly in two main categories: non-adaptive (e.g., \cite{macii2012,vejdan2017}) and adaptive (e.g., \cite{kamwa2014,serna2015}). On the one hand, non-adaptive algorithms have fixed parameters so that they generally provide good performance only in a limited set of operating conditions. On the other hand, adaptive solutions are much more flexible and, in principle, allow for good performance on a larger set of conditions. For instance, the well-known Interpolated DFT (IpDFT) algorithms enhance the DFT performance by introducing an adaptive compensation of the phasor estimate based on an estimate of the frequency deviation from nominal \cite{belega2013}. Similarly, by including the harmonic components in a state-space model, a Kalman filter can provide not only better off-nominal behavior than a single DFT but also improve its off-nominal harmonic rejection \cite{kamwa2014}. Finally, the so-called polynomial phase-locked loop Taylor-Fourier filters use an adaptive phase polynomial to center the Taylor expansion around the actual operating condition \cite{serna2015}, thus improving the approximation with respect to the plain Taylor-Fourier non-adaptive method \cite{serna2007}. However, adaptive systems are generally more computationally demanding. More seriously, convergence issues may appear in some operating conditions if this is neglected at the design stage. 

The main contribution of this work is to present the design of a novel adaptive dynamic phasor estimation algorithm. The proposed method offers a notable overall performance, a high harmonic rejection capability, and convergence guarantees in wide operating conditions that are specified at the design stage. It is based on a self-adaptive strategy working around a prototype differentiator filter bank (DFB) which allows to relax the design requirements of the filters and therefore leads to an enhanced performance. Convergence conditions are included in the design of the prototype system with the same convex semi-infinite programming (CSIP) framework as that used in \cite{messina2017}. Although the tool used for the design of the filters is the same as that in \cite{messina2017}, it should be emphasized that the algorithms are quite different. In fact, note that the method in \cite{messina2017} is a fixed filtering algorithm (i.e., non-adaptive).

\section{Prior Model and Parameter Shifting} \label{sec:model_and_demod}

For convenience of presentation, the signal phasor or synchrophasor (see definitions in \cite{ieeestd2011}) model is first presented in the continuous-time domain. 
Let $X(t) = a_{X}(t) e^{j \phi_{X}(t)}$, $t \in \R$, be the signal phasor, where $a_{X}(t) > 0$ is its (instantaneous) amplitude and $\phi_{X}(t)$ its phase. Then, around a particular fixed time, say the $r$-th reporting time $t_r$, we can write
\begin{equation} \label{eq:signal_phasor_ct} X(t_r+\tau) = X(t_r) e^{b(\tau)} e^{j \phi(\tau)},\ \tau \in \mathcal{T} \triangleq [-\delta, \delta], \end{equation}
where $b(\tau) = \ln a(\tau)$ is the log-amplitude modulation function, $a(\tau) = a_{X}(t_r+\tau)/a_{X}(t_r)$ is the amplitude modulation function, $\phi(\tau) =  \phi_{X}(t_r+\tau) - \phi_{X}(t_r)$ is the phase modulation function and $\delta > 0$ is a parameter that controls the width of the interval over which the signal phasor is to be approximated. In order to introduce the signal phasor prior model, the second-order log-amplitude and phase prior polynomials are defined as follows:
\begin{align} \label{eq:ref_pols} b_\r(\tau) & = -\sigma_\r \tau - \frac{1}{2} \gamma_\r \tau^2, \ \ \phi_\r(\tau) = \omega_\r \tau + \frac{1}{2} \alpha_\r \tau^2, \end{align}
where $\sigma_\r, \gamma_\r, \omega_\r, \alpha_\r$ are, respectively, the prior damping, rate of change of damping (ROCOD), frequency and rate of change of frequency (ROCOF). 
For instance, during a fault condition, a damped sinusoidal model is appropriate for describing the synchrophasor \cite{khodaparast2015} and damping gives important information about the instantaneous stability of the power system \cite{serna2013}. Moreover, second order information provided by the ROCOD may be valuable in order to predict critical grid operating conditions sooner. Finally, the signal phasor prior model may be written as $X_\r(\tau) = e^{b_\r(\tau)} e^{j \phi_\r(\tau)}$, $\tau \in \mathcal{T}$. Notice that this model is related to different proposals introduced in previous works. For instance, in \cite{serna2015} a Taylor polynomial is used to approximate the dynamic phasor with respect to an adaptive phase prior polynomial, but no adaptive prior model for the amplitude is proposed. On the other hand, Prony's method in \cite{serna2013} corresponds to an adaptive signal phasor model \mbox{$X(t_r + \tau) = X(t_r) e^{-\sigma \tau} e^{j \omega \tau}$}, $\tau \in \mathcal{T}$, where $\sigma$ and $\omega$ are the damping and frequency parameters, respectively. Therefore, $X_\r(\tau)$ gives a second-order generalization of this expression. Finally, in \cite{serna2007}, a Taylor expansion is proposed for the dynamic signal phasor. Here, instead, the approximations are made separately for amplitude and phase.


In what follows, the \textcolor{black}{parameter shifting} process is defined and analyzed. Let
\begin{equation} \label{eq:demod_Z_ct} Z(\tau) = X(t_r+\tau) e^{-b_\r(\tau)} e^{-j \phi_\r(\tau)} = a_{Z}(\tau) e^{j \phi_{Z}(\tau)}, \end{equation}
be the parameter shifted signal phasor, where its amplitude and phase are, respectively, ${a_{Z}(\tau) = a_{X}(t_r) e^{b(\tau) - b_\r(\tau)}}$ and ${\phi_{Z}(\tau) = \phi_{X}(t_r) + \phi(\tau) - \phi_\r(\tau)}$. 
Consider now the effect of this transformation at the reporting time $t_r$ (i.e., at $\tau = 0$). It is clear that for the phase, frequency (\textcolor{black}{i.e., the first time derivative of phase}) and ROCOF (\textcolor{black}{i.e., the second time derivative of phase}), the following relations hold:
\begin{align} \label{eq:phase_demod} \phi_{Z}(0) & = \phi_{X}(t_r) + \phi(0) - \phi_\r(0) =  \phi_{X}(t_r), \\ \label{eq:freq_demod} \omega_{Z}(0) & = \phi_{Z}'(0) = \phi'(0) - \phi_\r'(0) = \omega_{X}(t_r) - \omega_\r, \\ \label{eq:rocof_demod} \alpha_{Z}(0) & = \phi_{Z}''(0) = \phi''(0) - \phi_\r''(0) = \alpha_{X}(t_r) - \alpha_\r. \end{align}
On the other hand, for amplitude, damping (\textcolor{black}{i.e., minus the first time derivative of log-amplitude}) and ROCOD (\textcolor{black}{i.e., minus the second time derivative of log-amplitude}) one has, respectively,
\begin{align} \label{eq:amp_demod} a_{Z}(0) & = a_{X}(t_r) e^{b(0) - b_\r(0)} = a_{X}(t_r), \\ \label{eq:damping_demod} \sigma_{Z}(0) & = - \frac{a_{Z}'(0)}{a_{Z}(0)} = \sigma_{X}(t_r) - \sigma_\r, \\ \label{eq:rocod_demod} \gamma_{Z}(0) & = \sigma_{Z}'(0) = \gamma_{X}(t_r) - \gamma_\r. \end{align}
In summary, the amplitude and phase of $Z(\tau)$ at $\tau = 0$ are the same as those of $X(t_r)$, while the frequency, damping, ROCOF, and ROCOD are shifted by the prior parameters. This means that if the prior parameters match the ones of $X(t_r)$, the frequency, damping, ROCOF and ROCOD of $Z(\tau)$ will be zero at $\tau = 0$. In general, \textcolor{black}{if the match is not perfect but the prior parameters are not too different from the actual values}, the magnitude of the parameters of $Z(\tau)$ at $\tau = 0$ will be reduced with respect to those of $X(t_r)$. Therefore, the signal $\{ Z(\tau) \}_{\tau \in \mathcal{T}}$ is expected to be smoother than the signal ${\{ X(t_r+\tau) \}_{\tau \in \mathcal{T}}}$ if $\delta$ is sufficiently small.

\section{Self-Adaptive Contractive Algorithm} \label{sec:sadfb} 

\color{black}

\begin{figure}[ht]
	\centering \includegraphics[width = 0.94\columnwidth]{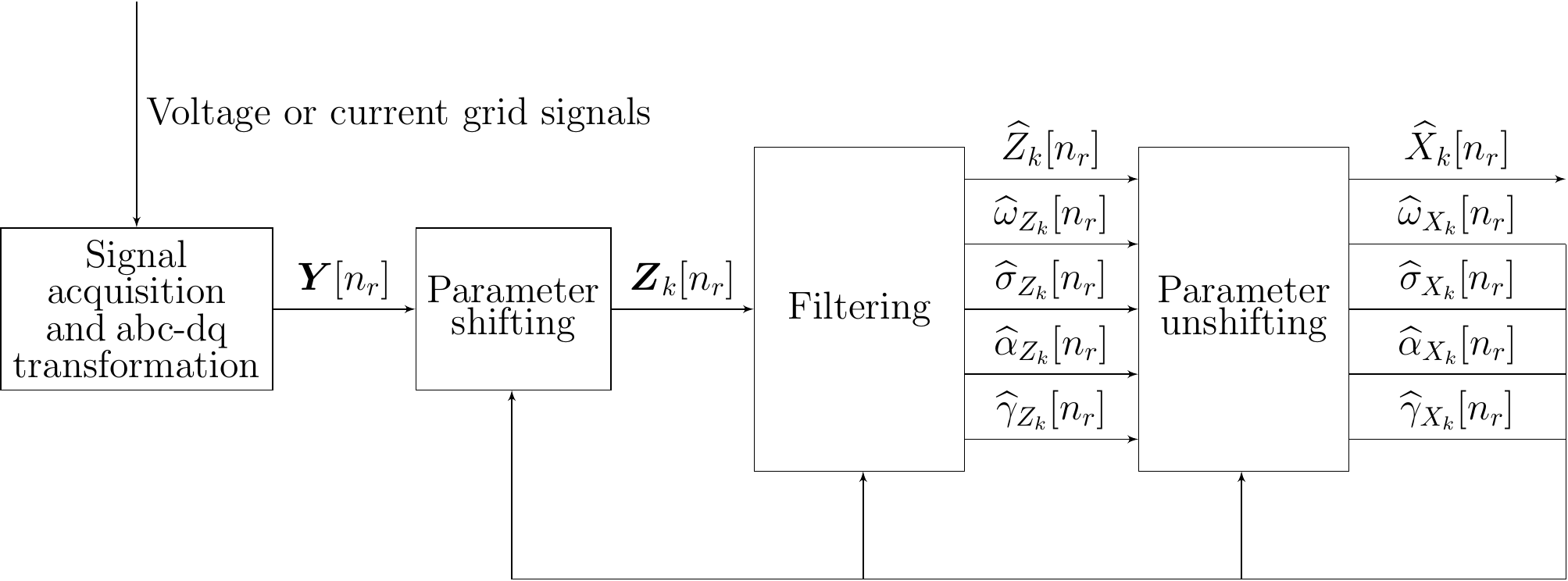}
	\centering \caption{SAC algorithm high-level block diagram.}
	\label{fig:sa_dfb}
\end{figure}

The self-adaptive contractive (SAC) algorithm proposed here is based on three stages: parameter shifting, filtering performed by an DFB, and parameter unshifting. In order to perform the system adaptation, the frequency, damping, ROCOF and ROCOD estimates of the signal phasor are fed back into the algorithm. This information is used to perform a parameter shifting in order to attract the input signal phasor to nominal conditions and, if necessary, to perform an adaptation of the filters coefficients in order to match their zeros with the off-nominal harmonic frequencies. After filtering the parameter shifted signal, a parameter unshifting of the estimates is performed to obtain the ones corresponding to the original signal phasor. The algorithm is iterative so it will produce estimates depending on an iteration index $k$, being the reported estimates the ones corresponding to the final iteration. The proposed method is shown schematically in Fig. \ref{fig:sa_dfb} and summarized as Algorithm \ref{algo:sac} at the end of this section for convenience and reference. In the following, a detailed description of each stage of the algorithm is provided.

\subsection{Input Signal Vector} \label{sec:sac_input_signal}

Consider the discrete-time signal phasor $X[n]$, $n \in \Z$. From \eqref{eq:signal_phasor_ct}, within the measurement window at the $r$-th reporting time, $X[n]$ takes the general form $X[n_r+m] = X[n_r] e^{b[m]} e^{j \phi[m]}$, $m \in \mathcal{W}$, where $n_r = t_r/T$ is assumed to be an integer, $b[m]$ and $\phi[m]$ are, respectively, the discrete-time log-amplitude and phase modulation functions, and \mbox{$\mathcal{W} = \{ -N/2, \ldots, N/2 \}$} is the window index set, being $N+1$ the window length, which is assumed to be odd. 
The input signal to the SAC algorithm is a complex baseband signal obtained from a real three-phase signal sampled with a sampling period $T$ and by performing an abc-dq transformation. As in \cite{messina2017}, the input signal is denoted as $Y[n] = X[n] + I[n]$, where $I[n]$ is a general disturbance signal which includes the harmonics, the negative-sequence component, noise, etc. The samples around each reporting time index $n_r$ are collected in the input signal vector \mbox{$\fg{Y}[n_r] = \left[ Y[n_r-N/2], \ldots, Y[n_r], \ldots, Y[n_r+N/2] \right]^T$}.

\subsection{Parameter Shifting} \label{sec:sac_demod}

From \eqref{eq:ref_pols}, the discrete-time phase prior polynomial at iteration $k$ is defined as follows
\begin{equation} \label{eq:phase_pol} \phi_{\r,k}[m] = \omega_{\r,k} T \, m + \frac{1}{2} \alpha_{\r,k} T^2 \, m^2, \qquad m \in \mathcal{W}, \end{equation}
where $\omega_{\r,k}$ and $\alpha_{\r,k}$ are, respectively, the prior frequency and ROCOF parameters at iteration $k$. These parameters are initialized using the previously reported estimates as ${\omega_{\r,0} = \widehat{\omega}_{X}[n_{r-1}] + \Delta t \, \widehat{\alpha}_{X}[n_{r-1}]}$ and \mbox{$\alpha_{\r,0} = \widehat{\alpha}_{X}[n_{r-1}]$}, where $\Delta t = 1/F_s$, being $F_s$ the reporting rate of the PMU. For the first window of samples, we use the nominal values of the parameters for initialization. Starting from these values, $\omega_{\r,k}$ and $\alpha_{\r,k}$ are adapted to match the signal phase dynamics at the current reporting time index $n_r$. Similarly, the log-amplitude polynomial is defined as follows: 
\begin{equation} \label{eq:log_amp_fun} b_{\r,k}[m] = -\sigma_{\r,k} T \, m - \frac{1}{2} \gamma_{\r,k} T^2 \, m^2, \qquad m \in \mathcal{W}, \end{equation}
where $\sigma_{\r,k}$ and $\gamma_{\r,k}$ are the prior damping and ROCOD parameters at iteration $k$. The initialization of these parameters is completely analogous to that of the parameters $\omega_{\r,k}$ and $\alpha_{\r,k}$. The adaptation scheme for the parameters $\omega_{\r,k}$, $\alpha_{\r,k}$, $\sigma_{\r,k}$ and $\gamma_{\r,k}$ is presented in Section \ref{sec:adap_crit}.

Now, the phasor prior matrix $\mat{D}_{\r,k} \in \C^{(N+1) \times (N+1)}$ is defined as a diagonal matrix with elements $(\mat{D}_{\r,k})_{m,m} = e^{b_{\r,k}[m]} \; e^{j \phi_{\r,k}[m]}$, 
where $m \in \mathcal{W}$. The parameter shifted input signal vector is then obtained as follows: 
\begin{equation} \label{eq:demod_phasor} \fg{Z}_k[n_r] = \mat{D}_{\r,k}^{-1} \; \fg{Y}[n_r]. \end{equation}
Notice that \eqref{eq:demod_phasor} is the discrete-time equivalent of \eqref{eq:demod_Z_ct}.

\subsection{Filtering} \label{sec:sac_filtering}

The estimates (respectively, the parameter shifted signal phasor estimate and its derivatives, frequency, damping, ROCOF and ROCOD) $\widehat{Z}_k[n_r]$, $\widehat{Z}'_k[n_r]$, $\widehat{Z}''_k[n_r]$, $\widehat{\omega}_{{Z}_k}[n_r]$,  $\widehat{\sigma}_{{Z}_k}[n_r]$, $\widehat{\alpha}_{{Z}_k}[n_r]$ and $\widehat{\gamma}_{Z_k}[n_r]$ are obtained by filtering $\fg{Z}_k[n_r]$ with the current DFB, which is represented by the linear-phase coefficient vectors $\f{a}_0 \in \R^{N/2+1}$, $\f{a}_1 \in \R^{N/2}$ and $\f{a}_2 \in \R^{N/2+1}$ \cite{messina2017}. The expressions are shown in \eqref{eq:z_hat} at the bottom of the next page (the rationale behind these computations can be found in \cite{serna2007}). For simplicity of notation, three constant matrices are defined, \mbox{$\mat{Q}_0 \in \R^{(N/2+1) \times (N+1)}$}, \mbox{$\mat{Q}_1 \in \R^{N/2 \times (N+1)}$} and \mbox{$\mat{Q}_2 \in \R^{(N/2+1) \times (N+1)}$}, to express the impulse response of the linear filters in terms of $\f{a}_0$, $\f{a}_1$ and $\f{a}_2$, respectively. The expressions of these matrices are omitted for space reasons.


\begin{figure*}[!hpb]
	\footnotesize
	\vspace*{4pt}
	\hrulefill

	\begin{align} \label{eq:z_hat} \widehat{Z}_k[n_r] & = \f{a}_0^T \mat{Q}_0 \fg{Z}_k[n_r], \qquad  && \widehat{Z}'_k[n_r] = \f{a}_1^T \mat{Q}_1 \fg{Z}_k[n_r], \qquad \qquad \qquad \qquad \widehat{Z}_k''[n_r] = \f{a}_2^T \mat{Q}_2 \fg{Z}_k[n_r], \nonumber \\ \widehat{\omega}_{Z_k}[n_r] & = \Im \left\{ \frac{\widehat{Z}'_k[n_r]}{\widehat{Z}_k[n_r]} \right\}, \qquad && \widehat{\alpha}_{Z_k}[n_r] = \Im \left\{ \frac{\widehat{Z}''_k[n_r]}{\widehat{Z}_k[n_r]} \right\} - 2 \Re \left\{ \frac{\widehat{Z}'_k[n_r]}{\widehat{Z}_k[n_r]} \right\} \Im \left\{ \frac{\widehat{Z}'_k[n_r]}{\widehat{Z}_k[n_r]} \right\}, \\ \widehat{\sigma}_{Z_k}[n_r] & = - \Re \left\{\frac{\widehat{Z}'_k[n_r]}{\widehat{Z}_k[n_r]} \right\}, \qquad && \widehat{\gamma}_{Z_k}[n_r] = - \Re \left\{ \frac{\widehat{Z}''_k[n_r]}{\widehat{Z}_k[n_r]} \right\} + \left[\Re \left\{\frac{\widehat{Z}'_k[n_r]}{\widehat{Z}_k[n_r]} \right\} \right]^2 - \left[ \Im \left\{ \frac{\widehat{Z}'_k[n_r]}{\widehat{Z}_k[n_r]} \right\} \right]^2 \nonumber. \end{align}

\end{figure*}




\subsection{Parameter Unshifting} \label{sec:comp}


The parameter unshifting stage is a transformation required to counteract the effects of the parameter shifting stage to obtain the desired parameter estimates. Note that, as in Section \ref{sec:model_and_demod}, the parameter unshifted estimates are denoted analogously to the parameter shifted ones by replacing $Z$ with $X$. From \eqref{eq:phase_demod} and \eqref{eq:amp_demod}, the transformation required for the phasor is trivial: \mbox{$\widehat{X}_k[n_r] = \widehat{Z}_k[n_r]$}. For the other parameters, the strategy is more complex for the sake of enhanced performance and robustness. Note that in an actual power system the range of values of the parameters is limited. However, some transient scenarios may produce very large misleading parameter estimates within a short time window. This observation leads to the following strategy to address the issue, which will be exemplified with the frequency estimate, but it applies in an exactly analogous manner to the other parameters. If \mbox{$|\widehat{\omega}_{Z_k}[n_r] + \omega_{\r,k} | \le \Delta \omega_{\max}$}, where $\Delta \omega_{\max}$ is an upper frequency threshold specified by the algorithm designer, a normal regime operation is considered. In such case, using the relation \eqref{eq:freq_demod}, the transformation is simply \mbox{$\widehat{\omega}_{X_k}[n_r] = \widehat{\omega}_{Z_k}[n_r] + \omega_{\r,k}$}. Otherwise, the estimate $\widehat{\omega}_{Z_k}[n_r]$ is disregarded and the transformation reduces to using the prior value: \mbox{$\widehat{\omega}_{X_k}[n_r] = \omega_{\r,k}$}. Note that to determine the parameter unshifting process completely we need the additional upper thresholds of damping, ROCOF and ROCOD (respectively, $\Delta \sigma_{\max}, \Delta \alpha_{\max}, \Delta \gamma_{\max}$).

\subsection{Prior Parameters Adaptation} \label{sec:adap_crit}

\begin{figure}[htbp]
	\centering \includegraphics[width = 0.95\columnwidth]{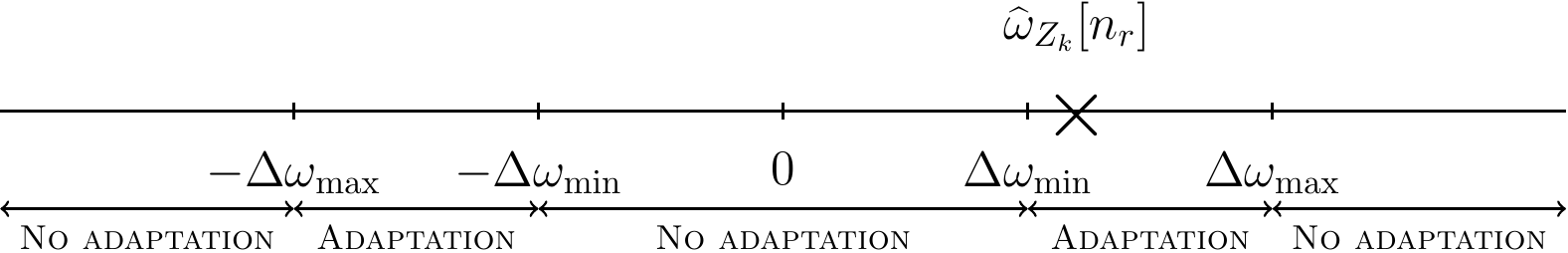}
	\centering \caption{Illustration of adaptation criteria for prior frequency.}
	\label{fig:adap_criteria}
\end{figure}

Note that at a particular reporting time $n_r$, there may be some parameters that require adaptation and others that do not. For instance, during the onset of a frequency ramp, only the ROCOF prior should be modified. In addition, a large and fast frequency change may indicate a phase step, in which case the algorithm should not modify wildly its prior parameter and filter coefficients but rather perform as prescribed by the prototype filters in that transient condition. Concretely, the algorithm should avoid very large changes in the prior parameters for the sake of reliability as well as very small changes for the sake of computational efficiency. With this purpose in mind, we introduce the algorithm lower thresholds: $\Delta \omega_{\min}$, $\Delta \alpha_{\min}$, $\Delta \sigma_{\min}$, and $\Delta \gamma_{\min}$. Now, the parameter adaptation set (i.e., the prior parameters for which the adaptation will be made at iteration $k$) \mbox{$P_k \subseteq \{ \omega_{\r,k}, \alpha_{\r,k}, \sigma_{\r,k}, \gamma_{\r,k} \}$} is defined according to whether the quantities $|\widehat{\omega}_{Z_k}[n_r]|, |\widehat{\alpha}_{Z_k}[n_r]|, |\widehat{\sigma}_{Z_k}[n_r]|$, and $|\widehat{\gamma}_{Z_k}[n_r]|$ belong to its corresponding ranges defined by the lower and upper algorithm thresholds. For example, \mbox{$\omega_{\r,k} \in P_k$} if and only if ${\Delta \omega_{\min} \le |\widehat{\omega}_{Z_k}[n_r]| \le \Delta \omega_{\max}}$. These criteria are illustrated in Fig. \ref{fig:adap_criteria} for the frequency. For each parameter in $P_k$, its value is updated according to the following simple rules: $\omega_{\r,k+1} = \widehat{\omega}_{X_k}[n_r]$, $\alpha_{\r,k+1} = \widehat{\alpha}_{X_k}[n_r]$, $\sigma_{\r,k+1} = \widehat{\sigma}_{X_k}[n_r]$, and $\gamma_{\r,k+1} = \widehat{\gamma}_{X_k}[n_r]$. The algorithm will then iterate until either $P_k = \emptyset$ or $k$ becomes greater than some predefined limit $k_{\max}$.





\subsection{Filters Coefficients Adaptation} \label{sec:sac_filters_adap}

In order to guarantee that the DFB behaves properly with large off-nominal and high total harmonic distortion levels, it is required to adapt the filters to have zeros as close as possible to the off-nominal harmonic frequencies. In addition, it is desirable that the filters have ideal gains at the fundamental frequency. It is possible to express these conditions as linear constraints in the filter coefficients $\f{a}_i$ and therefore they can be written in matrix form as \mbox{$\mat{G}_i \f{a}_i = \f{c}_i$}. The matrices $\mat{G}_i$ depend on $\omega_{h,l} = l (\omega_0 + \widehat{\omega}_{X_{k-1}}[n_r])$, \mbox{$l = 1, \ldots, M$}, the off-nominal harmonic frequency estimates available at iteration $k$ prior to filtering, where $\omega_0$ is the angular nominal power system frequency, and $M$ is the number of harmonics considered. The expressions of $\mat{G}_i$ are omitted for space reasons, while $\f{c}_0 = [1, 0, \ldots, 0]^T \in \R^{N/2+1}, \f{c}_1 = \f{0} \in \R^{N/2}, \f{c}_2 = \f{0} \in \R^{N/2+1}$.



On the other hand, the goal of the filters adaptation is to modify the filters coefficients as little as possible in order to preserve the prototype filter features, but also enforcing the aforementioned constraints. The $L^2[-\pi, \pi]$ squared distance between the adapted filters and the prototype filters are, by Parseval's theorem, 
given by ${\| \mat{P}_i^{1/2} (\f{a}_i - \widetilde{\f{a}}_i) \|^2}$, where $\mat{P}_0 = \diag(1, \frac{1}{2}, \ldots, \frac{1}{2}) = \mat{P}_2$, and $\mat{P}_1 = \diag(\frac{1}{2}, \ldots, \frac{1}{2})$, are diagonal matrices \cite{schafer2009}. 
Therefore, the adapted filters coefficients are obtained by solving the following linearly constrained least-squares optimization problem:
\begin{equation} \label{eq:lcls_problem_1}
\begin{aligned}
& \underset{\f{a}_i}{\min} & & \| \mat{P}_i^{1/2} (\f{a}_i - \widetilde{\f{a}}_i) \|^2 & \text{\textcolor{black}{subject to}}	& & \mat{G}_i \f{a}_i = \f{c}_i.
\end{aligned}
\end{equation}
The solution to this problem can be readily found by using the method of Lagrange multipliers, yielding
\begin{equation} \label{eq:lcls_solution_1} \f{a}_i = \widetilde{\f{a}_i} + \mat{P}_i^{-1} \mat{G}_i^T (\mat{G}_i \mat{P}_i^{-1} \mat{G}_i^T)^{-1} (\f{c}_i - \mat{G}_i \widetilde{\f{a}_i}). \end{equation}
Note that the adaptation of the filters coefficients will occur if and only if $\omega_{\r,k} \in P_k$.


\subsection{Computational Cost}

The online computational cost of each stage of the SAC algorithm for one iteration is as follows. The computation of the prior polynomials, the parameter shifting operation, and the filtering stage have a cost of $O(N)$ operations. The parameter unshifting and prior parameters adaptation take only $O(1)$ operations. Finally, the filters coefficients adaptation has a cost of $O(M^2 N)$ operations, where it is assumed that $N \gg M$. Therefore, the adaptation of the filters coefficients is clearly the dominant cost component of the algorithm. Thus, a proper selection of the frequency thresholds $\Delta \omega_{\min}$ and $\Delta \omega_{\max}$ should be made to minimize it as much as possible for a given desired performance in a practical setting. Note also that the overall cost of the algorithm is similar or lower than that of comparable self-adaptive solutions \cite{petri2014,serna2015}.

\begin{algorithm}[ht]
	\footnotesize
	\caption{SAC} \label{algo:sac}
	\begin{algorithmic}[*]
		\Input $\{Y[n]\}$, $\f{\widetilde{a}}_i$, $k_{\max}$, $M$, $\Delta \omega_{\min}$, $\Delta \omega_{\max}$, $\Delta \sigma_{\min}$, $\Delta \sigma_{\max}$, $\Delta \alpha_{\min}$, \\ \pushcode[0] $\Delta \alpha_{\max}$, $\Delta \gamma_{\min}$, $\Delta \gamma_{\max}$. 
		\Output $\{\widehat{X}[n_r]\}, \{\widehat{\omega}_{X}[n_r]\}$, $\{\widehat{\alpha}_{X}[n_r]\}$, $\{\widehat{\sigma}_{X}[n_r]\}$, $\{\widehat{\gamma}_{X}[n_r]\}$.

		\State Initialize the filters coefficients as $\f{a}_i = \widetilde{\f{a}}_i$ (see Section \ref{sec:pfbd}).
		\For {each reporting time $n_r$}

		\State {{\bf Initialization} (Sections \ref{sec:sac_input_signal} and \ref{sec:sac_demod})}
		\State Set $k : = 0$ and form the input signal phasor vector $\fg{Y}[n_r]$.
		\State Initialize $\omega_{\r,k}$, $\alpha_{\r,k}$, $\sigma_{\r,k}$, $\gamma_{\r,k}$.
		\State Compute $b_{\r,k}[m]$, $\phi_{\r,k}[m]$. 
		\State {{\bf \textcolor{black}{Parameter shifting}} (Section \ref{sec:sac_demod})}
		\State Compute $\fg{Z}_k[n_r]$.

		\State {{\bf{Filtering}} (Section \ref{sec:sac_filtering})}
		\State Find $\widehat{Z}_k[n_r]$, $\widehat{\omega}_{Z_k}[n_r]$, $\widehat{\alpha}_{Z_k}[n_r]$, $\widehat{\sigma}_{Z_k}[n_r]$, $\widehat{\gamma}_{Z_k}[n_r]$.

		\State {{\bf \textcolor{black}{Parameter unshifting}} (Section \ref{sec:comp})}
		\State Find $\widehat{X}_k[n_r]$, $\widehat{\omega}_{X_k}[n_r]$, $\widehat{\alpha}_{X_k}[n_r]$, $\widehat{\sigma}_{X_k}[n_r]$ and $\widehat{\gamma}_{X_k}[n_r]$. 





		\State {{\bf Define the Parameter Adaptation Set} (Section \ref{sec:adap_crit})}

		\State Define $P_k$.

		\While {$P_k \ne\emptyset$ and $k < k_{\max}$}

		\State {{\bf Update of Prior Parameters} (Section \ref{sec:adap_crit})}

		\State Update each parameter in $P_k$. Set $k := k+1$.

		\State {{\bf Computation of Prior Polynomials} (Section \ref{sec:sac_demod})}
		\State Compute $b_{\r,k}[m]$ and $\phi_{\r,k}[m]$.

		\State {{\bf \textcolor{black}{Parameter shifting}} (Section \ref{sec:sac_demod})}
		\State Compute $\fg{Z}_k[n_r]$.

		\State {{\bf Filters Adapation} (Section \ref{sec:sac_filters_adap}) }
		\If {$\omega_{\r,k} \in P_k$}
		\State Compute $\mat{G}_i$ and update $\f{a}_i$.
		\EndIf

		\State {{\bf{Filtering}} (Section \ref{sec:sac_filtering})}
		\State Compute $\widehat{Z}_k[n_r]$, $\widehat{\omega}_{Z_k}[n_r]$, $\widehat{\alpha}_{Z_k}[n_r]$, $\widehat{\sigma}_{Z_k}[n_r]$ and $\widehat{\gamma}_{Z_k}[n_r]$.

		\State {{\bf\textcolor{black}{Parameter unshifting}} (Section \ref{sec:comp})}
		\State Compute $\widehat{X}_k[n_r]$, $\widehat{\omega}_{X_k}[n_r]$, $\widehat{\alpha}_{X_k}[n_r]$, $\widehat{\sigma}_{X_k}[n_r]$ and $\widehat{\gamma}_{X_k}[n_r]$.

		\State {{\bf Update of Parameter Adaptation Set} (Section \ref{sec:adap_crit})}

		\State Update $P_k$.

		\EndWhile

		\State {\bf Estimates Reporting}
		\State Set $\widehat{X}[n_r] := \widehat{X}_k[n_r]$, $\widehat{\omega}_{X}[n_r] := \widehat{\omega}_{X_k}[n_r]$,  $\widehat{\alpha}_{X}[n_r] := \widehat{\alpha}_{X_k}[n_r]$, \\ \pushcode[0] $\widehat{\sigma}_{X}[n_r] := \widehat{\sigma}_{X_k}[n_r]$, $\widehat{\gamma}_{X}[n_r] := \widehat{\gamma}_{X_k}[n_r]$.

		\EndFor
	\end{algorithmic}
\end{algorithm}


\section{Prototype Differentiator Filter Bank Design} \label{sec:pfbd}


The DFB consists of a parallel architecture of three linear-phase filters to estimate the signal phasor and its first two derivatives \cite{messina2017}. The order of the filters are all assumed to be equal to $N$, and the expressions for their amplitude responses are given by $\widetilde{A}_i(\omega T) = \f{g}_i^T(\omega T) \; \widetilde{\f{a}}_i,\ i = 0,1,2,$
where $\f{g}_0(\omega T) = [1, \cos(\omega T), \ldots, \cos(\omega T N/2)]^T$, $\f{g}_1(\omega T) = [\sin(\omega T), \ldots, \sin(\omega T N/2)]^T$ and $\f{g}_2(\omega T) = \f{g}_0(\omega T)$. 
The analysis and design of an DFB structure using the CSIP framework is thoroughly presented in \cite{messina2017}, so the reader is referred to that paper for the details of this approach. Here, we only discuss the differences that arise in the design of the DFB when it is to be used in the SAC algorithm. As will be shown, the self-adaptation scheme relaxes most of the performance constraints of the problem. Moreover, in many cases, the constraints imply that, under such cases, the different errors, which are the total vector error (TVE), frequency error (FE), rate of change of frequency error (RFE), damping error (DE) and rate of change of damping error (RDE), converge to zero (see Appendix), which is a much stronger result than a performance bound on them as in the formulation of \cite{messina2017}. 

\subsection{Cost Functions} \label{sec:cost_functions}

The cost functions are chosen in order to obtain minimum norm filters, yielding
\begin{align} f_i(\widetilde{\f{a}}_i) = \frac{1}{2\pi} \|\widetilde{A}_i\|^2 = \widetilde{\f{a}}_i^T \mat{P}_i \widetilde{\f{a}}_i, \qquad i = 0,1,2, \end{align}
It can be proved that this choice minimizes the error contribution of a small model mismatch between the prior phasor and the signal phasor and, in fact, it is interesting to note that the DFT and Taylor-Fourier filters can also be interpreted as minimum norm filters. However, note that the design constraints and the structure of these algorithms are very different.

\subsection{Interference Rejection Constraints}

As discussed in Section \ref{sec:sac_filters_adap}, in off-nominal frequency conditions, the filters adaptation strategy forces zeros at the corresponding frequencies. Therefore, only linear constraints on $\widetilde{\f{a}}_i$ to ensure zeros at nominal harmonics are required:
\begin{equation} \widetilde{A}_i(l \omega_0 T) = 0, \qquad i = 0,1,2, \qquad l = 1, \ldots, M. \end{equation}
This relaxation of the original harmonic rejection constraints is of critical importance. In fact, if one includes the joint effect of off-nominal frequency conditions and multiple high-level harmonics as constraints in the original CSIP filter design problem, the resulting optimization problems may result unfeasible \cite{messina2017}. 




\subsection{Contractive Constraints} \label{sec:contractive_constraints}


In steady-state off-nominal frequency conditions, the signal phasor around a fixed reporting time $n_r$ can be modeled as $X[n_r+m] = X[n_r] e^{j \omega_s T m}$, where $\omega_s \in \Omega_{\con}$. A sufficient condition for the convergence of $\widehat{\omega}_{X_k}[n_r]$ to $\omega_s$ as $k \to \infty$ is given by the following convex semi-infinite (CSI) contractive constraint on $\widetilde{\f{a}}_1$ (see Appendix for details)
\begin{equation} \label{eq:freq_contraction} \left| \omega_s - \frac{\widetilde{A}_1(\omega_s T)}{\widetilde{A}_0(\omega_s T)} \right| \le L_{\omega} |\omega_s|, \qquad \forall \, \omega_s \in \Omega_{\con}. \end{equation}
The set $\Omega_{\con}$ is called the contractive frequency set and is defined as $\Omega_{\con} = [-\omega_{\con} \, T, \omega_{\con} \, T]$, where $\omega_{\con}$ is the frequency contractive range (in rad/s). Also, $L_{\omega} \in (0,1)$ is the frequency contraction parameter, which controls the rate of convergence of the frequency estimate. This constraint, imposed on the design of the filter $\widetilde{A}_1$ given $\widetilde{A}_0$, is convex. It is interesting to compare this constraint with the one obtained in \cite{messina2017}. Note that the bound $L_{\omega} |\omega_s|$ is greater than $2 \pi \FE_{\STA}$ for all frequencies $f_s$ such that $|f_s| > \FE_{\STA}/ L_{\omega}$, which in general is in the order of a few mHz. This shows that a relaxation of the original constraint is achieved for significant off-nominal frequencies.

Similar contractive constraints are posed for damping, ROCOF, and ROCOD, by considering, respectively, the following signal phasor models:
\begin{align} X[n_r+m] & = X[n_r] e^{-\sigma_s T m},  && \sigma_s \in \Sigma_{\con}, \\ X[n_r+m] & = X[n_r] e^{j \frac{1}{2} \alpha_s T^2 m^2}, && \alpha_s \in A_{\con}, \\ X[n_r+m] & = X[n_r] e^{-\frac{1}{2} \gamma_s T^2 m^2} , && \gamma_s \in \Gamma_{\con}. \end{align}
Note that, once again, the constraints are completely specified by their contractive ranges ($\sigma_{\con}, \alpha_{\con}, \gamma_{\con}$) and contraction parameters ($L_{\sigma}, L_{\alpha}, L_{\gamma}$). The constraint expressions are omitted for the sake of brevity.

\subsection{Other Constraints}

A contractive constraint in frequency, damping, ROCOF and ROCOD is sufficient to guarantee good performance in amplitude and phase modulation conditions, so no additional constraints are introduced for this scenario. However, it should be noted that in these cases there is a small model mismatch between the prior phasor and the true signal phasor, even with perfect prior parameters, but its order is $O(T^3)$. Moreover, as discussed in Section \ref{sec:cost_functions}, the impact of this mismatch is minimized by the choice of the cost functions.




On the other hand, as explained in Section \ref{sec:adap_crit}, during a step transient, the SAC algorithm should behave like a fixed DFB if the thresholds are correctly chosen. Therefore, the required constraints are the same as those presented in \cite{messina2017}.

\begin{table*}
	\caption{Worst-case error metrics for the different tests of Sections \ref{sec:numerical_stationary}, \ref{sec:sim_dynamics} and \ref{sec:sim_faults}}
	\label{tab:test_results}
	\centering
	\begin{tabular}{cc|c|c|c|c|c|c|c|}
		\cline{3-9}
		& & AE (pu) & PE (rad) & TVE (\%) & FE (Hz) & DE (1/s) & RFE (Hz/s) & RDE (1/s$^2$) \\ \hline
		\multicolumn{1}{|c|}{\multirow{2}{*}{A1}} & SAC \textcolor{black}{algorithm}      & $8.9415 \cdot 10^{-5}$                & $9.0944\cdot 10^{-5}$                    & 0.0096                    & 0.0012                   & 0.0078                    & 0.1688 & 1.0641                  \\ \cline{2-9}
		\multicolumn{1}{|c|}{}                           & IEEE Std. \textcolor{black}{algorithm} & 0.2193               & $8.2336\cdot 10^{-5}$                     & 21.9288                   & 0.0034                   & 0.1423                    & 14.7719 & 598.5252                   \\ \hline
		\multicolumn{1}{|c|}{\multirow{2}{*}{A2}}     & SAC  \textcolor{black}{algorithm}     & $9.7480 \cdot 10^{-5}$               & $1.0019 \cdot 10^{-4}$                    & 0.0102                    & 0.0072                   & 0.0078                    & 0.1744 & 1.0670                  \\ \cline{2-9}
		\multicolumn{1}{|c|}{}                           & IEEE Std. \textcolor{black}{algorithm} & 0.2231               & 0.0026                     & 22.3070                   & 0.1782                   & 6.5213                    & 554.8150  & $2.5687 \cdot 10^{4}$                 \\ \hline
		\multicolumn{1}{|c|}{\multirow{2}{*}{A3}}     & SAC \textcolor{black}{algorithm}      & $9.4300\cdot 10^{-5}$                & $1.0835\cdot 10^{-4}$                    & 0.0112                   & 0.0084                   & 0.0144                    & 0.1555   & 1.0552                 \\ \cline{2-9}
		\multicolumn{1}{|c|}{}                           & IEEE Std. \textcolor{black}{algorithm}& 0.3031               & 0.0487                    & 30.3112                   & 7.6028                   & 321.4530                    & $1.7683\cdot 10^{4}$    & $1.1325\cdot 10^{6}$               \\ \hline
		\multicolumn{1}{|c|}{\multirow{2}{*}{B1}}     & SAC \textcolor{black}{algorithm}      & $6.3672\cdot 10^{-5}$               & $6.5001\cdot 10^{-5}$                    & 0.0068                   & 0.0010                   & 0.0055                    & 0.1004     & 0.4069              \\ \cline{2-9}
		\multicolumn{1}{|c|}{}                           & IEEE Std. \textcolor{black}{algorithm} & 0.2187               & $3.5572\cdot10^{-4}$                    & 21.8652                   & 0.0036                   & 0.1383                     & 9.6899      & 450.7530             \\ \hline
		\multicolumn{1}{|c|}{\multirow{2}{*}{B2}}     & SAC \textcolor{black}{algorithm}      & $7.5743\cdot10^{-5}$               & $7.4745\cdot10^{-5}$                    & 0.0084                   & 0.0016                   & 0.0090                    & 0.2663       & 1.4563            \\ \cline{2-9}
		\multicolumn{1}{|c|}{}                           & IEEE Std. \textcolor{black}{algorithm} & $7.7686\cdot10^{-4}$               & $7.8009\cdot10^{-4}$                    & 0.1144                   & 0.0032                   & 0.0199                    & 6.7630        & 45.1783           \\ \hline
		\multicolumn{1}{|c|}{\multirow{2}{*}{B3}}     & SAC \textcolor{black}{algorithm}      & $1.8363\cdot10^{-4}$               & $1.7879\cdot10^{-4}$                    & 0.0233                   & 0.0121                   & 0.0806                   & 1.2611         & 8.6851          \\ \cline{2-9}
		\multicolumn{1}{|c|}{}                           & IEEE Std. \textcolor{black}{algorithm} & 0.0047               & 0.0047                    & 0.6981                   & 0.0261                   & 0.1603                    & 8.2548          & 55.2348         \\ \hline
		\multicolumn{1}{|c|}{\multirow{2}{*}{C1}}     & SAC \textcolor{black}{algorithm}    & $6.5510\cdot10^{-5}$ & $6.4767\cdot10^{-5}$ & 0.0070 & $8.8005\cdot10^{-4}$ & 0.0067 & 0.1372 & 1.0653 \\ \cline{2-9}
		\multicolumn{1}{|c|}{}                           & IEEE Std. \textcolor{black}{algorithm} & 0.0026 & $4.7742\cdot10^{-4}$ & 0.2501 & 0.0023 & 0.0152 & 6.7818 & 48.3326 \\ \hline
		\multicolumn{1}{|c|}{\multirow{2}{*}{C2}}     & SAC \textcolor{black}{algorithm}  & $7.5912\cdot10^{-5}$ & $6.6228\cdot10^{-5}$ & 0.0074 & $8.5263\cdot10^{-4}$ & 0.0057 & 0.1290 & 0.7502  \\ \cline{2-9}
		\multicolumn{1}{|c|}{}                           & IEEE Std. \textcolor{black}{algorithm} & $7.8080\cdot10^{-4}$ & 0.0010 & 0.1303 & 0.0023 & 0.0136 & 8.5739 & 50.0218 \\ \hline
		\multicolumn{1}{|c|}{\multirow{2}{*}{C3}}     & SAC \textcolor{black}{algorithm}      & $7.3535\cdot10^{-5}$ & $6.6000\cdot10^{-5}$ & 0.0075 & $8.3744\cdot10^{-4}$ & 0.0074 & 0.1367 & 1.2403 \\ \cline{2-9}
		\multicolumn{1}{|c|}{}                           & IEEE Std. \textcolor{black}{algorithm} & 0.0011 & $7.0928\cdot10^{-4}$ & 0.1253 & 0.0025 & 0.0161 & 7.5543 & 52.7143         \\ \hline
	\end{tabular}
\end{table*}

\section{Numerical Results} \label{sec:sim}

For the sake of a better assessment of the results, the algorithm was implemented in a sliding-window fashion. 
The following parameter values are used in all simulations. The nominal frequency is set to $f_0 = 60$ Hz and the sampling period is $T = 1/(32 \cdot f_0) \approx 520.83$ $\mu$s. The reporting rate is set to $F_s = 60$ frames-per-second (fps). After a parameter tuning process, the thresholds are defined to be as follows: $\Delta \omega_{\min} = 2 \pi \cdot 10^{-3}$ rad/s, $\Delta \omega_{\max} = 2\pi \cdot 15$ rad/s, $\Delta \sigma_{\min} = 6 \cdot 10^{-3}$ 1/s, $\Delta \sigma_{\max} = 4$ 1/s, $\Delta \alpha_{\min} = 2 \pi \cdot 0.1$ rad/s$^2$, $\Delta \alpha_{\max} = 2\pi \cdot 16$ rad/s$^2$, $\Delta \gamma_{\min} = 0.6$ 1/s$^2$, and $\Delta \gamma_{\max} = 110$ 1/s$^2$. The number of iterations is limited to $k_{\max} = 5$. Due to aliasing effects, only $M = 11$ harmonics are considered instead of 50 as in the IEEE Std. In any case, one should consider that in practice the anti-aliasing filter limits the harmonic content of the signal fed to the algorithm. To properly account for noise, note that the signal-to-noise ratio (SNR) was measured at the distribution level to be around 60 dB \cite{zhan2015}, while SNR at the transmission level is higher. However, it should be noted that this value was obtained with a sampling rate of 1024 samples per cycle. Assuming that noise is white its power may be reduced approximately by the factor $1024/32 = 32$ by using a proper decimator filter. Thus, we use $\SNR = 75$ dB, which is the expected SNR at the sampling rate considered.

The prototype filters length is set to two nominal cycles, which is particularly well suited for protection and control applications (i.e., for use in a P class PMU). Parameter values for their design are $\omega_{\con} = 2 \pi \cdot 15$ rad/s, $\alpha_{\con} = 2 \pi \cdot 16$ rad/s$^2$, $\sigma_{\con} = 4$ 1/s, $\gamma_{\con} = 110$ 1/s$^2$, $L_{\omega} = L_{\sigma} = 0.3$ and $L_{\alpha} = L_{\gamma} = 0.9$. 
\textcolor{black}{The contractive ranges were chosen to ensure convergence over the most stringent range of parameters of the benchmark signals considered in the IEEE Std. However, $\omega_{\con}$ is chosen so that the algorithm has a wide frequency range basin of attraction, motivated by other works \cite{kamwa2014,zhan2018}.} Note that the contractive constraints on the second-order variables are weaker, which gives looser constraints for the prototype CSIP filters design problem and therefore allows to obtain a smaller value for their norms. This translates to better performance in noisy conditions, which is a critical issue for ROCOF and ROCOD estimates. For the step constraints, the uniform scaling parameters are set to $\rho_i = 0.5$ (see \cite{messina2017}). 

%

%

The worst-case errors of the SAC algorithm for the different tests performed are summarized in Table \ref{tab:test_results}. In addition, as a benchmark, the same values are reported for the method proposed in the IEEE Std. for a P class PMU (from now on simply the IEEE Std. algorithm) \cite{ieeestd2011,ieeestd2014}. The step test results are presented in Table \ref{tab:step_results} as they involve different metrics of performance. For the sake of completeness, damping and ROCOD estimates were incorporated to the IEEE Std. algorithm using central finite differences analogous to the ones used for frequency and ROCOF estimation but applied to the estimate of the log-amplitude signal. 


%
%
%
%
%


\subsection{Stationary Tests} \label{sec:numerical_stationary}



\color{black}

\begin{table*}
	\caption{Amplitude and phase step performance results}
	\label{tab:step_results}
	\centering
	\begin{tabular}{cc|c|c|c|c|c|c|}
		\cline{3-8}
		&           & Overshoot (\%) & $t_{\res,\TVE}$ (s) & $t_{\res,\FE}$ (s) & $t_{\res,\DE}$ (s) & $t_{\res,\RFE}$ (s) & \textcolor{black}{$t_{\res,\RDE}$} (s) \\ \hline
		\multicolumn{1}{|c|}{\multirow{2}{*}{Amplitude}} & SAC \textcolor{black}{algorithm}    & 1.5877               & 0.0141                    & 0                   & 0.0339                  & 0                    & 0.0385                    \\ \cline{2-8}
		\multicolumn{1}{|c|}{}                           & IEEE Std. \textcolor{black}{algorithm} & 0.0282               & 0.0271                    & 0                   & 0.0349                   & $\infty$                    & $\infty$                    \\ \hline
		\multicolumn{1}{|c|}{\multirow{2}{*}{Phase}}     & SAC \textcolor{black}{algorithm}      & 0.7111               & 0.0161                    & 0.0339                   & 0.0339                   & 0.0401                    & 0.0380                    \\ \cline{2-8}
		\multicolumn{1}{|c|}{}                           & IEEE Std. \textcolor{black}{algorithm} & 0.0191               & 0.0307                    & 0.0349                   & 0.0359 & $\infty$                   & $\infty$                                         \\ \hline
	\end{tabular}
\end{table*}

\subsubsection{Off-nominal frequency}

The system is tested with a sinusoidal input signal whose frequency is swept in the off-nominal range defined by the contractive set $\Omega_{\con}$ with steps of 0.1 Hz (\textcolor{black}{referred to as test} A1). The results show a good performance of the SAC algorithm over the entire contractive frequency range. 
Moreover, it is interesting to observe that the values of $2 \pi \cdot \FE$ and DE are comparable, as well as those of $2 \pi \cdot \RFE$ and RDE. This could be useful to set tentative limits for these quantities.

\subsubsection{Harmonic distortion}

The harmonic distortion rejection test is considered under two different scenarios, jointly with an off-nominal frequency condition in the entire frequency contractive range. In the first scenario, each harmonic is introduced separately with an amplitude equal to $1 \%$ of the fundamental \textcolor{black}{(test A2)}. In the second scenario, all harmonic amplitudes are simultaneously set to $10 \%$ of the fundamental \textcolor{black}{(test A3)}. Although this is an unrealistic condition, it is nevertheless useful to verify the SAC algorithm robustness to high harmonic distortion levels. In addition, it is interesting to note that in all of the above tests, the number of iterations $k$ was only nonzero at the initialization. Therefore, in steady-state conditions, the cost of the algorithm is virtually the same as that of a non-adaptive DFB. 

\subsection{Dynamic Tests} \label{sec:sim_dynamics}


\subsubsection{Frequency Ramp}

\color{black}

A frequency ramp test is performed starting at $t = 1$ s and ranging from 45 to 75 Hz with a ramp rate of 1 Hz/s (\textcolor{black}{test} B1). Maximum errors on this time interval (discarding exclusion intervals \cite{ieeestd2014}) are presented in Table \ref{tab:test_results}. Again, due to the wide frequency range covered in the test, the performance difference of the methods is quite large.


\color{black}


%

\subsubsection{Amplitude and Phase Modulations}

A joint amplitude and phase modulation test is performed using a modulation frequency of 2 Hz and modulation factors of 0.1 (test B2). The results of the SAC algorithm shown in Table \ref{tab:test_results} indicate that the model mismatch impact is not severe for such low-frequency modulations. In fact, once again, no adaptations occur after the initialization. On the other hand, the same test but for a 5 Hz modulation frequency is also considered (test B3). In this case, the model mismatch impact on the SAC algorithm is much more noticeable. In fact, adaptations occur on more than $67 \%$ of the times, with 0.94 average iterations. Nevertheless, results are well within the IEEE Std. established limits \cite{ieeestd2014}. 

\subsubsection{Amplitude and Phase Steps}

\color{black}

Firstly, an amplitude step test with a step size of $0.1$ pu is performed. It should be noted that the DE and RDE response times have been defined by using the rough empirical correspondences \mbox{$\DE \sim 2 \pi \cdot \FE$} and \mbox{$\RDE \sim 2 \pi \cdot \RFE$} in order to assess their speed of response. Response times are denoted as $t_{\res,\TVE}, t_{\res,\FE}$, etc. Secondly, a phase step test with a step size of $\pi/18$ rad is considered. The results in Table \ref{tab:step_results} show that the SAC algorithm adaptation scheme works successfully under fast changes. In fact, the performance is compliant with the prototype specifications. Note that the IEEE Std. algorithm RFE and RDE response times are reported as infinity since these error metrics are always greater than the limits established in the IEEE Std.


\color{black}

\subsection{Faults} \label{sec:sim_faults}



In all fault conditions, the performance is reported in Table \ref{tab:test_results} after the response time allowed by the IEEE Std. (i.e., $2/f_0 \approx 33.3$ ms for AE, PE and TVE, $4.5/f_0 = 75$ ms for DE and FE, and $6/f_0 = 100$ ms for RFE and RDE) in a step test.

\subsubsection{Small Power System} 
\begin{figure}[ht]
	\includegraphics[width=1\columnwidth]{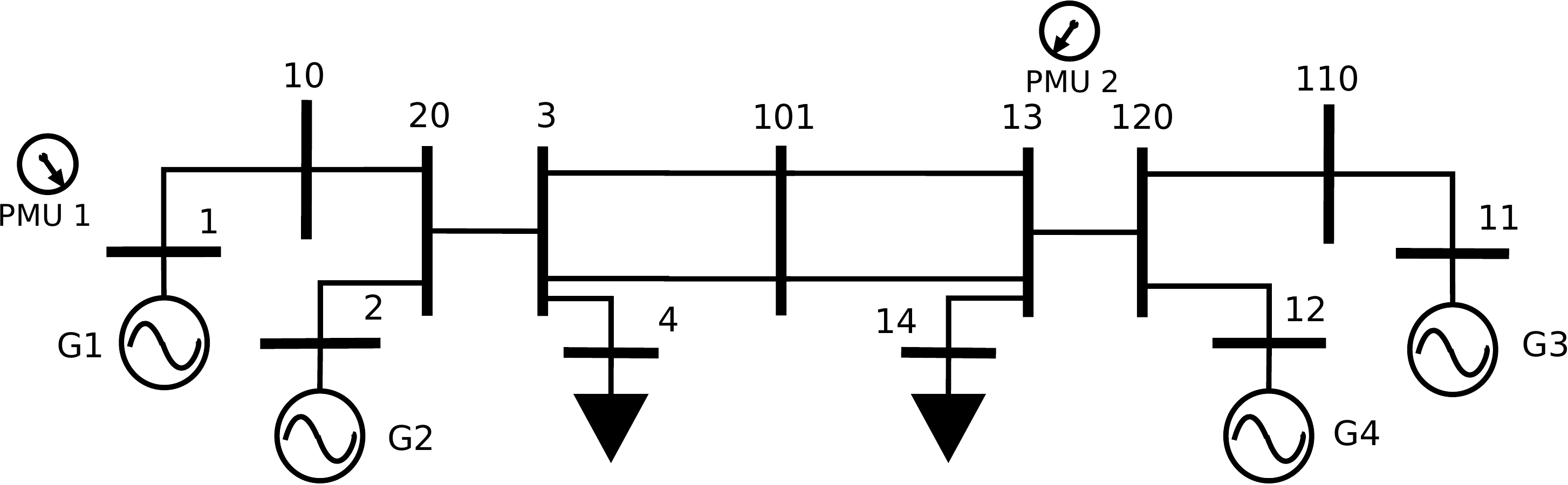}
	\caption{Single line diagram of the small power system.}
	\label{fig:test_system}
\end{figure}

\begin{figure}[ht]
	\centering		
	\begin{tabular}{cc}

		\hspace{-5mm} 
		{
			\includegraphics[width = 0.48 \columnwidth]{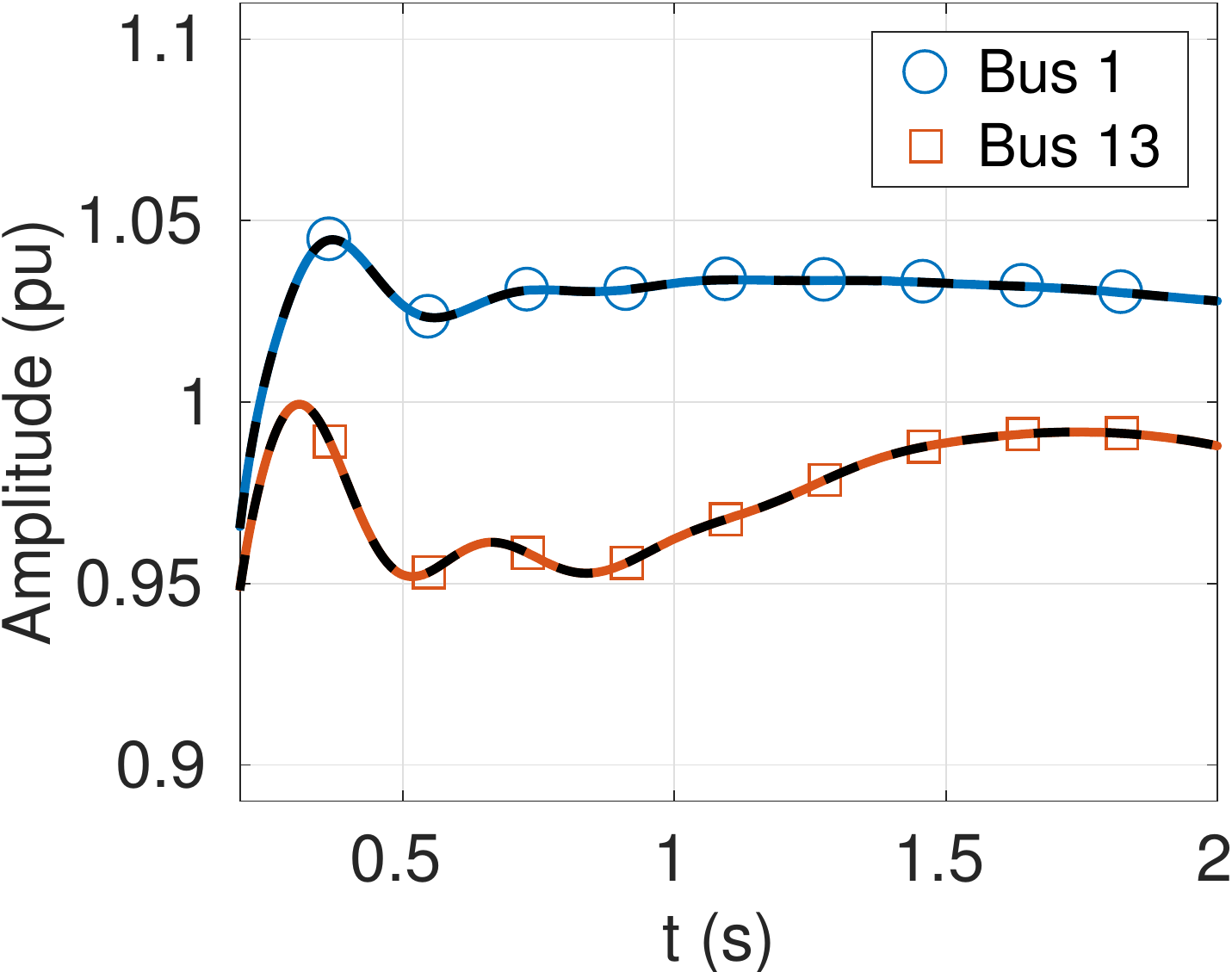}} &
		\hspace{-5mm} 
		{
			\includegraphics[width = 0.45 \columnwidth]{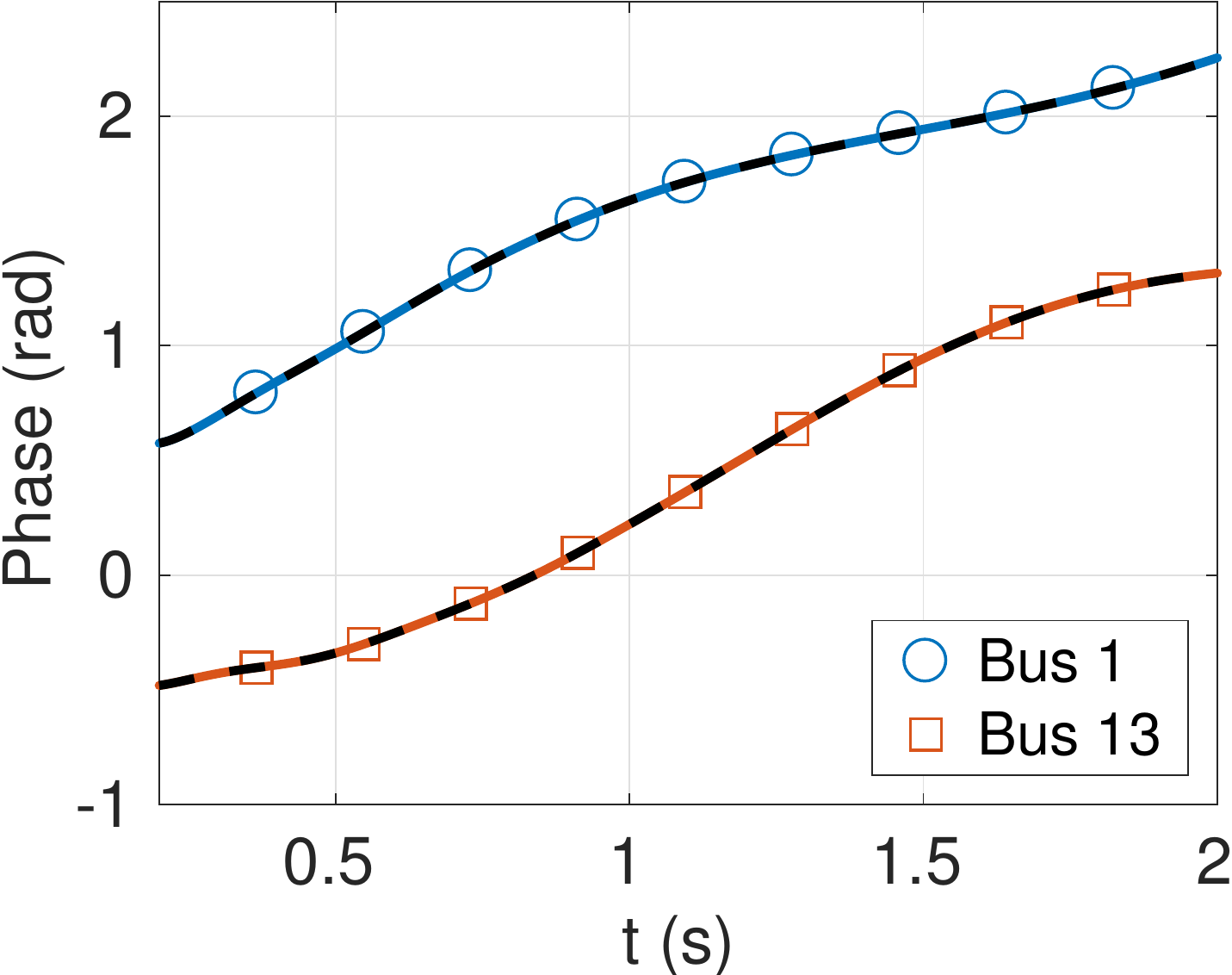}} \\
		\hspace{-5mm} 
		{
			\includegraphics[width = 0.46 \columnwidth]{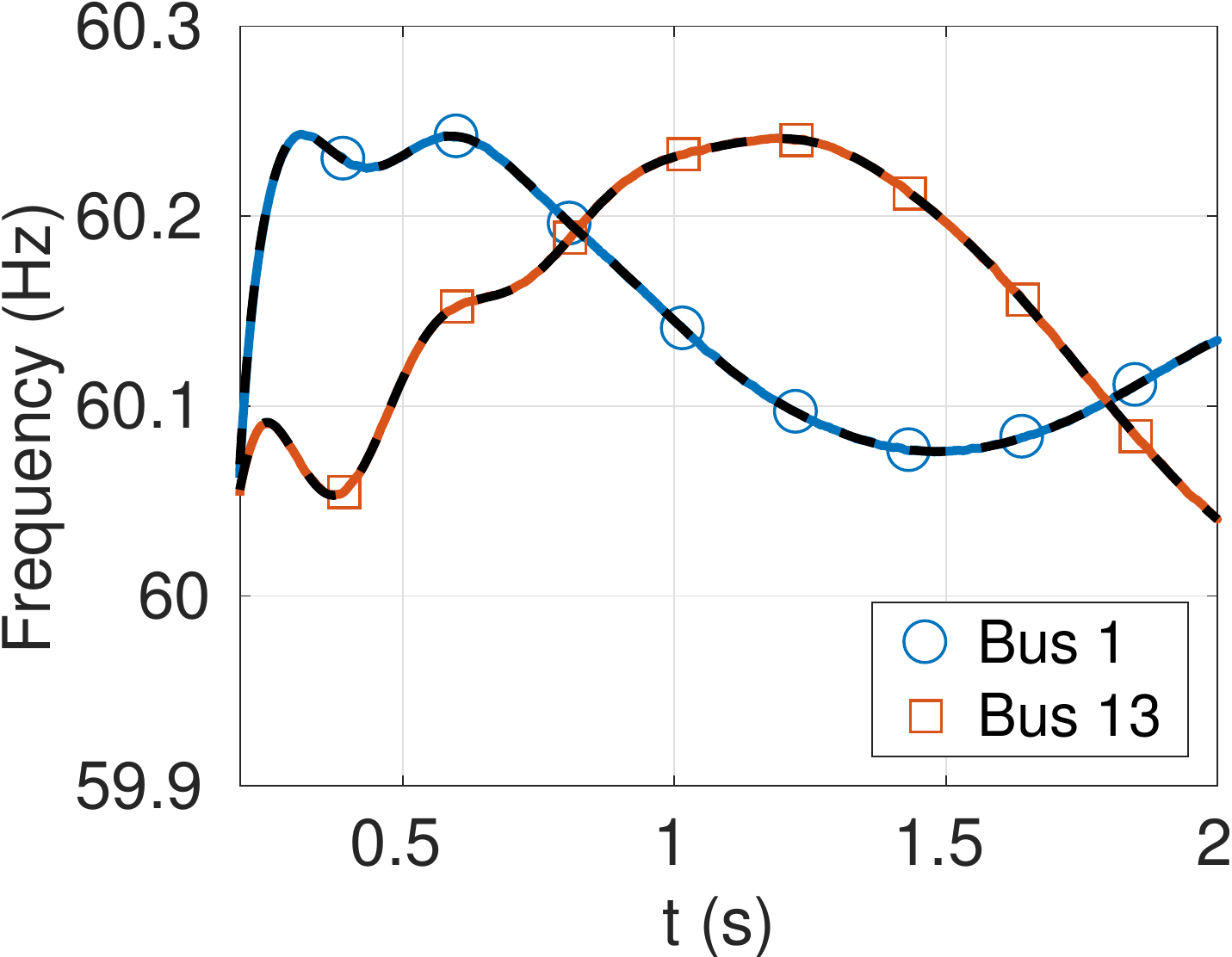}} &
		\hspace{-5mm} 
		{
			\includegraphics[width = 0.48 \columnwidth]{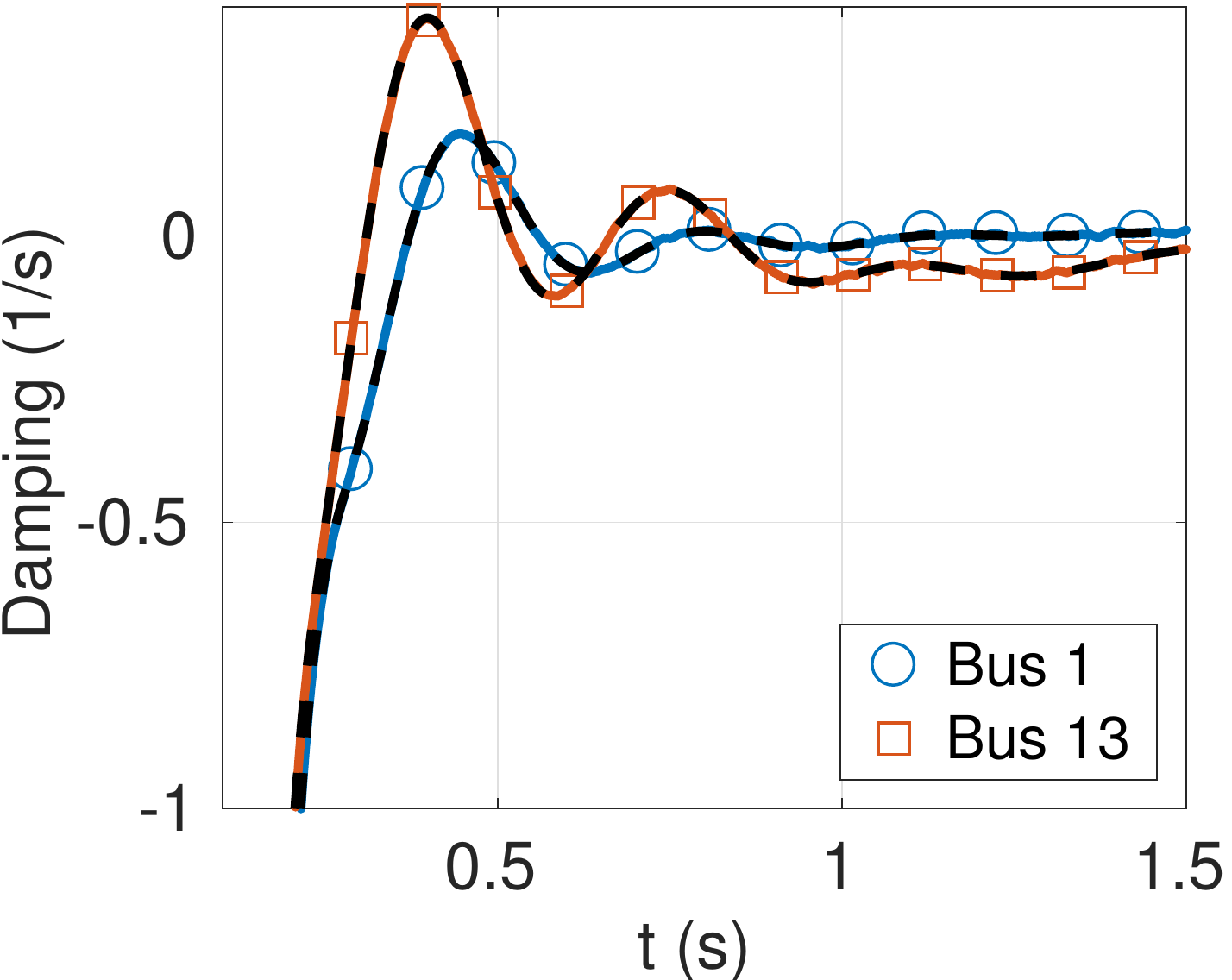}} \\
		\hspace{-5mm} 
		{
			\includegraphics[width = 0.48 \columnwidth]{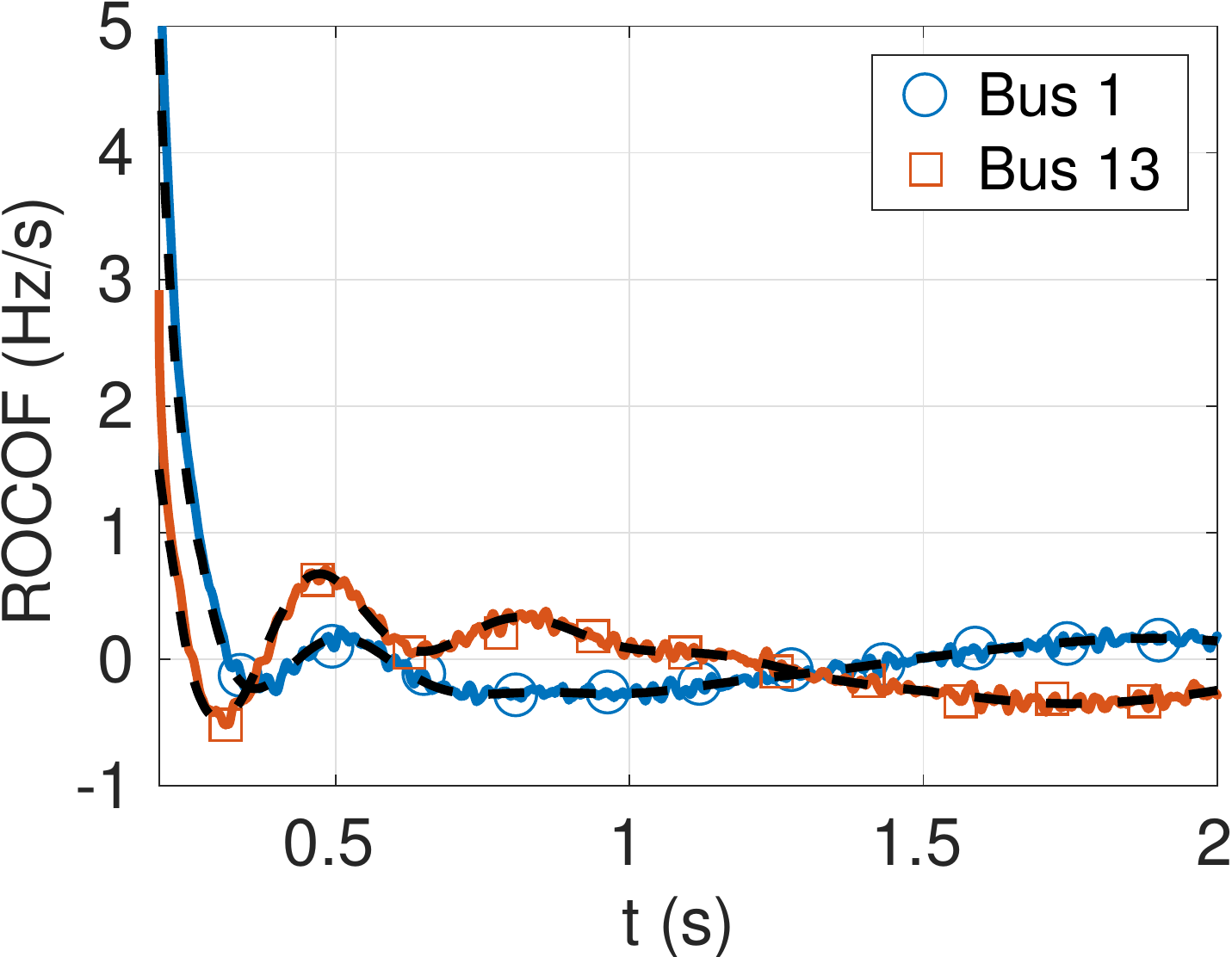}} &
		\hspace{-5mm} 
		{
			\includegraphics[width = 0.47 \columnwidth]{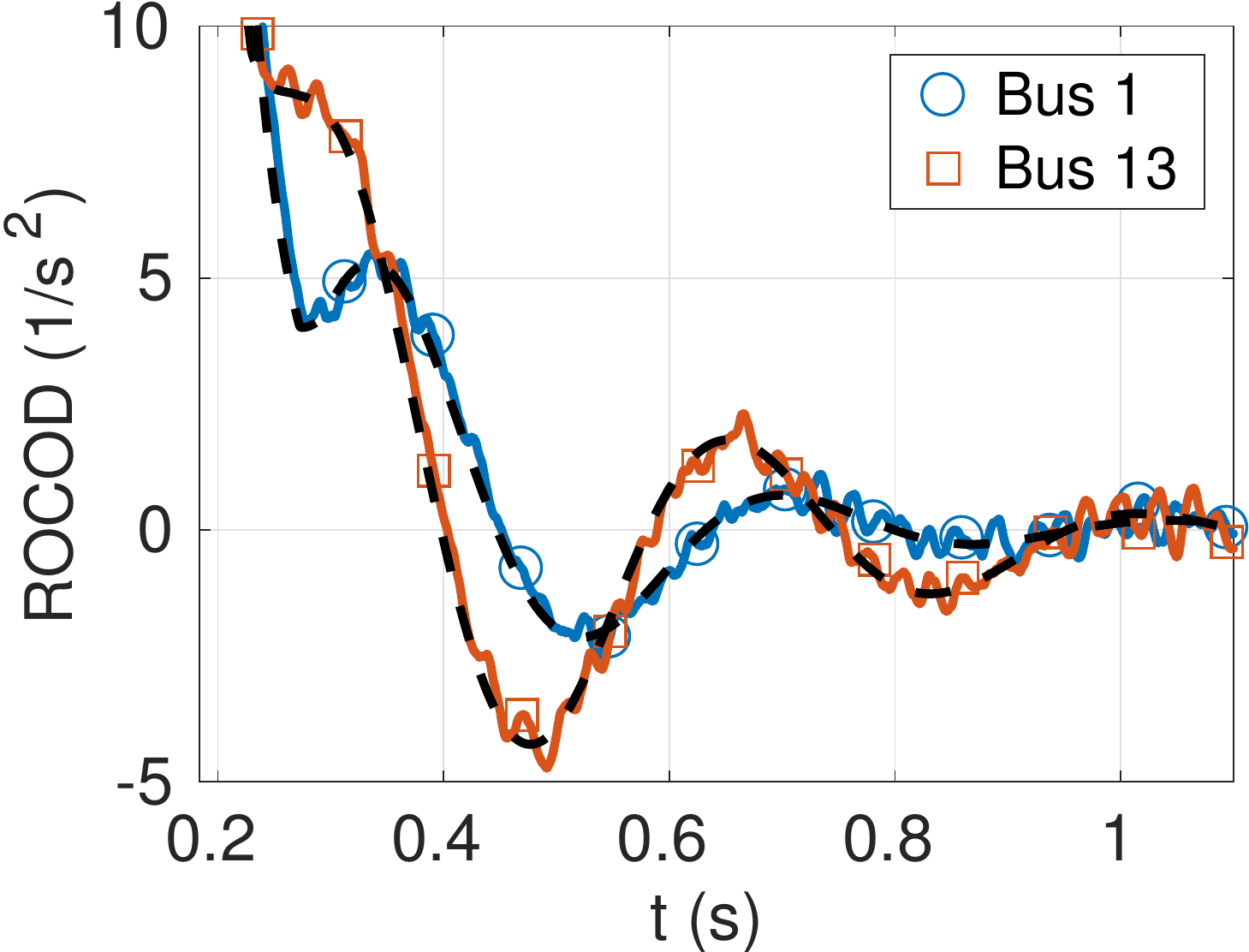}} \\

	\end{tabular}
	\caption{Three-phase fault results for the small power system. Theoretical values of the parameters are shown as dashed black lines.} \label{fig:fault_results}
\end{figure}

The two-area system of Fig. \ref{fig:test_system} is considered \cite{Klein1991}, where it is assumed that two PMUs are installed at buses 1 and 13. The simulation is performed with the Power System Toolbox (PST) \cite{Chow1992} (using the \textit{d2asbeg.m} file with default parameters) and numerical results are reported for the PMU at bus 1. Two situations are considered. First, a multiple loss of lines is applied at \mbox{$t=0.1$ s} between the buses 3 and 101 (test C1). Results in Table \ref{tab:test_results} show that in such severe case the algorithm will allow for a smooth transition between the two grid topologies. Second, a three-phase fault is applied at the same time and between the same buses, and it is cleared after 5 cycles (test C2). The evolution of the estimates of all the parameters is shown in Fig. \ref{fig:fault_results}, showing a close agreement between the true values and the estimates after a fast transient.






\subsubsection{Large Power System} A three-phase fault between buses 6 and 7 is simulated for the well-known IEEE 50-machine 149-bus system, and measurements at bus 6 are assessed (test C3). In this case, we observed that the dynamics are significantly more pronounced but correctly tracked by the SAC algorithm.




%

\section{Conclusion} \label{sec:con}

A novel algorithm for enhanced dynamic phasor estimation with a self-adaptive capability has been presented to cope with the stringent performance requirements found in different applications of PMUs measurements. There are several interesting features that make this algorithm a powerful and flexible tool. Firstly, it has a guaranteed convergence in a prescribed set of conditions that can be defined at the design stage. Secondly, it offers great harmonic rejection even with large frequency deviations. Thirdly, it offers a simple control on the trade-off between accuracy, speed and computational cost. In addition, it is proposed to estimate (and report) damping and ROCOD, which are more physically meaningful than plain amplitude derivatives. Thus, the knowledge of these quantities could lead to the design of novel amplitude control schemes. Simulations of the IEEE Std. tests under realistic noisy conditions as well as different faults have shown that remarkably accurate and fast estimates can be obtained with this approach. These results show that algorithm is promising for use in both transmission and distribution networks.



\appendix[Convergence Analysis] \label{sec:conv}


For brevity, only the evolution of frequency is considered. However, the other quantities are treated in an exactly analogous manner. Assuming for simplicity that the adaptation is performed always, the prior frequency at iteration $k+1$ is simply $\omega_{\r,k+1} = \omega_{\r,k} + \widehat{\omega}_{Z_k}[n_r]$. To derive the convergence conditions, consider the signal phasor model ${X[n_r+m]} = X[n_r] e^{j \omega_s T m}$, where $\omega_s \in \Omega_{\con}$. Neglecting the influence on the other parameters for simplicity (i.e., considering a one-dimensional error flow), it is found that $e_{\omega,k} = \omega_{s} - \omega_{\r,k}$ satisfies the recursive relation
\begin{equation} e_{\omega,k+1} = e_{\omega,k} - \widehat{\omega}_{Z_{k+1}}[n_r] = e_{\omega,k} - \frac{A_1(e_{\omega,k} T)}{A_0(e_{\omega,k} T)} \triangleq f_{\omega}(e_{\omega,k}). \end{equation}
It is clear that $f_{\omega}$ has a fixed point at the origin (i.e., \mbox{$f_{\omega}(0) = 0$}) since \mbox{$A_1(0) = 0$} for any \mbox{$\f{a}_1 \in \R^{N/2}$}. Therefore, convergence of the prior frequency $\omega_{\r,k}$ to $\omega_s$, for any $\omega_s \in \Omega_{\con}$, can be guaranteed if the following contractive constraint holds:
\begin{equation} \label{eq:freq_convergence_condition} |f_{\omega}(\omega_s)| = \left| \omega_s - \frac{A_1(\omega_s T)}{A_0(\omega_s T)} \right| \le L_{\omega} |\omega_s|, \qquad \forall \, \omega_s \in \Omega_{\con}. \end{equation}
This is actually an application of the celebrated contraction mapping theorem, which guarantees both uniqueness of the fixed point zero and convergence of the fixed-point iteration $e_{\omega,k+1} = f_{\omega}(e_{\omega,k})$ for all the desired conditions \cite{kolmogorov1999}. By design, from \eqref{eq:freq_contraction}, condition \eqref{eq:freq_convergence_condition} holds exactly for the prototype filters, that is, for the function $\widetilde{f}_{\omega}(\omega_s) = \omega_s - \widetilde{A}_1(\omega_s T)/\widetilde{A}_0(\omega_s T)$. In addition, the filters adaptation strategy \eqref{eq:lcls_solution_1} changes the filter amplitude responses mainly around harmonic frequencies. Thus, the approximation $f_{\omega}(\omega_s) \approx \widetilde{f}_{\omega}(\omega_s)$ for all $\omega_s \in \Omega_{\con}$ is reasonable for our analysis. This implies that \mbox{$Z_k[n_r+m] \to X[n_r]$} for all \mbox{$m \in \mathcal{W}$} as $k \to \infty$, that is, $\fg{Z}_k[n_r]$ converges to a constant signal. Therefore, the DFB ideal gain nominal frequency conditions $A_0(0) = 1$ and $A_1(0) = 0$ guarantee that $\widehat{\omega}_{X_k}[n_r] \to \omega_s$ as $k \to \infty$.


%







\ifCLASSOPTIONcaptionsoff
  \newpage
\fi




\bibliographystyle{IEEEtran}
\bibliography{paper}

\begin{thebibliography}{10}
\providecommand{\url}[1]{#1}
\csname url@samestyle\endcsname
\providecommand{\newblock}{\relax}
\providecommand{\bibinfo}[2]{#2}
\providecommand{\BIBentrySTDinterwordspacing}{\spaceskip=0pt\relax}
\providecommand{\BIBentryALTinterwordstretchfactor}{4}
\providecommand{\BIBentryALTinterwordspacing}{\spaceskip=\fontdimen2\font plus
\BIBentryALTinterwordstretchfactor\fontdimen3\font minus
  \fontdimen4\font\relax}
\providecommand{\BIBforeignlanguage}[2]{{%
\expandafter\ifx\csname l@#1\endcsname\relax
\typeout{** WARNING: IEEEtran.bst: No hyphenation pattern has been}%
\typeout{** loaded for the language `#1'. Using the pattern for}%
\typeout{** the default language instead.}%
\else
\language=\csname l@#1\endcsname
\fi
#2}}
\providecommand{\BIBdecl}{\relax}
\BIBdecl

\bibitem{ipakchi2009}
A.~Ipakchi and F.~Albuyeh, ``{Grid of the Future},'' \emph{IEEE Power and
  Energy Magazine}, vol.~7, no.~2, pp. 52--62, March 2009.

\bibitem{tielens2016}
P.~Tielens and D.~V. Hertem, ``The relevance of inertia in power systems,''
  \emph{Renewable and Sustainable Energy Reviews}, vol.~55, pp. 999 -- 1009,
  2016.

\bibitem{aminifar2014}
F.~Aminifar, M.~Fotuhi-Firuzabad, A.~Safdarian, A.~Davoudi, and
  M.~Shahidehpour, ``{Synchrophasor Measurement Technology in Power Systems:
  Panorama and State-of-the-Art},'' \emph{IEEE Access}, vol.~2, pp. 1607--1628,
  2014.

\bibitem{ieeestd2011}
\emph{{IEEE Standard for Synchrophasor Measurements for Power Systems}}, IEEE
  Std. C37.118.1-2011 (Revision of IEEE Std C37.118-2005), Dec. 2011.

\bibitem{ieeestd2014}
\emph{{IEEE Standard for Synchrophasor Measurements for Power Systems --
  Amendment 1: Modification of Selected Performance Requirements}}, IEEE Std.
  C37.118.1a-2014 (Amendment to IEEE Std C37.118.1-2011), Apr. 2014.

\bibitem{phadke2017}
A.~Phadke and J.~Thorp, \emph{Synchronized Phasor Measurements and Their
  Applications}, ser. Power Electronics and Power Systems.\hskip 1em plus 0.5em
  minus 0.4em\relax Springer, 2017.

\bibitem{barchi2015}
G.~Barchi, D.~Fontanelli, D.~Macii, and D.~Petri, ``{On the Accuracy of Phasor
  Angle Measurements in Power Networks},'' \emph{IEEE Transactions on
  Instrumentation and Measurement}, vol.~64, no.~5, pp. 1129--1139, May 2015.

\bibitem{meier2017}
A.~von Meier, E.~Stewart, A.~McEachern, M.~Andersen, and L.~Mehrmanesh,
  ``{Precision Micro-Synchrophasors for Distribution Systems: A Summary of
  Applications},'' \emph{IEEE Transactions on Smart Grid}, vol.~8, no.~6, pp.
  2926--2936, Nov 2017.

\bibitem{roscoe2015}
A.~J. Roscoe, B.~Dickerson, and K.~E. Martin, ``{Filter Design Masks for
  C37.118.1a-Compliant Frequency-Tracking and Fixed-Filter M-Class Phasor
  Measurement Units},'' \emph{IEEE Transactions on Instrumentation and
  Measurement}, vol.~64, no.~8, pp. 2096--2107, Aug 2015.

\bibitem{kamwa2014}
I.~Kamwa, S.~R. Samantaray, and G.~Joos, ``{Wide Frequency Range Adaptive
  Phasor and Frequency PMU Algorithms},'' \emph{IEEE Transactions on Smart
  Grid}, vol.~5, no.~2, pp. 569--579, March 2014.

\bibitem{zhan2018}
L.~Zhan, Y.~Liu, and Y.~Liu, ``{A Clarke Transformation-Based DFT Phasor and
  Frequency Algorithm for Wide Frequency Range},'' \emph{IEEE Transactions on
  Smart Grid}, vol.~9, no.~1, pp. 67--77, Jan 2018.

\bibitem{macii2016}
D.~Macii, D.~Fontanelli, G.~Barchi, and D.~Petri, ``{Impact of Acquisition
  Wideband Noise on Synchrophasor Measurements: A Design Perspective},''
  \emph{IEEE Transactions on Instrumentation and Measurement}, vol.~65, no.~10,
  pp. 2244--2253, Oct 2016.

\bibitem{macii2012}
D.~Macii, D.~Petri, and A.~Zorat, ``{Accuracy Analysis and Enhancement of
  DFT-Based Synchrophasor Estimators in Off-Nominal Conditions},'' \emph{IEEE
  Transactions on Instrumentation and Measurement}, vol.~61, no.~10, pp.
  2653--2664, Oct. 2012.

\bibitem{vejdan2017}
S.~Vejdan, M.~Sanaye-Pasand, and O.~P. Malik, ``{Accurate Dynamic Phasor
  Estimation Based on the Signal Model Under Off-Nominal Frequency and
  Oscillations},'' \emph{IEEE Transactions on Smart Grid}, vol.~8, no.~2, pp.
  708--719, March 2017.

\bibitem{serna2015}
J.~A. de~la O~Serna, ``{Synchrophasor Measurement With Polynomial
  Phase-Locked-Loop Taylor-Fourier Filters},'' \emph{IEEE Transactions on
  Instrumentation and Measurement}, vol.~64, no.~2, pp. 328--337, Feb 2015.

\bibitem{belega2013}
D.~Belega and D.~Petri, ``{Accuracy Analysis of the Multicycle Synchrophasor
  Estimator Provided by the Interpolated DFT Algorithm},'' \emph{IEEE
  Transactions on Instrumentation and Measurement}, vol.~62, no.~5, pp.
  942--953, May 2013.

\bibitem{serna2007}
J.~A. de~la O~Serna, ``{Dynamic Phasor Estimates for Power System
  Oscillations},'' \emph{IEEE Transactions on Instrumentation and Measurement},
  vol.~56, no.~5, pp. 1648--1657, Oct. 2007.

\bibitem{messina2017}
F.~Messina, L.~{Rey Vega}, P.~Marchi, and C.~G. Galarza, ``{Optimal
  Differentiator Filter Banks for PMUs and Their Feasibility Limits},''
  \emph{IEEE Transactions on Instrumentation and Measurement}, vol.~66, no.~11,
  pp. 2948--2956, Nov 2017.

\bibitem{khodaparast2015}
J.~Khodaparast and M.~Khederzadeh, ``{Three-Phase Fault Detection During Power
  Swing by Transient Monitor},'' \emph{IEEE Transactions on Power Systems},
  vol.~30, no.~5, pp. 2558--2565, Sept 2015.

\bibitem{serna2013}
J.~A. de~la O~Serna, ``{Synchrophasor Estimation Using Prony's Method},''
  \emph{IEEE Transactions on Instrumentation and Measurement}, vol.~62, no.~8,
  pp. 2119--2128, Aug 2013.

\bibitem{schafer2009}
A.~V. Oppenheim and R.~W. Schafer, \emph{{Discrete-Time Signal Processing}},
  3rd~ed.\hskip 1em plus 0.5em minus 0.4em\relax Upper Saddle River, NJ:
  Prentice Hall Press, 2009.

\bibitem{petri2014}
D.~Petri, D.~Fontanelli, and D.~Macii, ``{A Frequency-Domain Algorithm for
  Dynamic Synchrophasor and Frequency Estimation},'' \emph{IEEE Transactions on
  Instrumentation and Measurement}, vol.~63, no.~10, pp. 2330--2340, Oct. 2014.

\bibitem{zhan2015}
L.~Zhan, Y.~Liu, J.~Culliss, J.~Zhao, and Y.~Liu, ``{Dynamic Single-Phase
  Synchronized Phase and Frequency Estimation at the Distribution Level},''
  \emph{IEEE Transactions on Smart Grid}, vol.~6, no.~4, pp. 2013--2022, July
  2015.

\bibitem{Klein1991}
M.~Klein, G.~J. Rogers, and P.~Kundur, ``{A Fundamental Study of Inter-Area
  Oscillations in Power Systems},'' \emph{IEEE Transactions on Power Systems},
  vol.~6, no.~3, pp. 914--921, Aug 1991.

\bibitem{Chow1992}
J.~H. Chow and K.~W. Cheung, ``{A Toolbox for Power System Dynamics and Control
  Engineering Education and Research},'' \emph{IEEE Transactions on Power
  Systems}, vol.~7, no.~4, pp. 1559--1564, Nov 1992.

\bibitem{kolmogorov1999}
A.~Kolmogorov and S.~Fomin, \emph{{Elements of the Theory of Functions and
  Functional Analysis}}.\hskip 1em plus 0.5em minus 0.4em\relax Mineola, NY:
  Dover, 1999.

\end{thebibliography}

\end{document}